\documentclass{article}
\usepackage{spconf,amsmath,graphicx}

\usepackage{color}
\usepackage{url}
\usepackage{enumitem}

\usepackage{caption}
\title{DEEP EXPOSURE FUSION WITH DEGHOSTING VIA HOMOGRAPHY ESTIMATION AND ATTENTION LEARNING}
\name{Sheng-Yeh Chen$\quad$Yung-Yu Chuang%\thanks{Thanks to XYZ agency for funding.}
}
\address{National Taiwan University}
\begin{document}
%\ninept
%
\maketitle
\tabcolsep=2pt

\definecolor{green}{rgb}{0, 0.6, 0} 

%%% Frequently used terms.
\newcommand{\etal}{et~al.}
\newcommand{\ie}{i.e.}
\newcommand{\eg}{e.g.}

\newcommand{\heading}[1]{\noindent\textbf{#1}}

\newcommand{\figref}[1]{Fig.~\ref{fig:#1}}% type "\figref{}" to reference figure
\newcommand{\tabref}[1]{Table~\ref{tab:#1}} % type "\tabref{}" to reference table
\newcommand{\itmref}[1]{[\ref{itm:#1}]}     % type "\itmref{}" to reference item
\newcommand{\eqnref}[1]{Equation~\ref{eq:#1}}
\newcommand{\secref}[1]{Section~\ref{sec:#1}}

\newcommand{\tb}[1]{\textbf{#1}}
\newcommand{\mb}[1]{\mathbf{#1}}
\newcommand{\bs}[1]{\boldsymbol{#1}}
\newcommand{\MR}[1]{\multirow{2}{*}{#1}}
\newcommand{\norm}[1]{\left\|#1\right\|}
\newcommand{\n}{\mbox{\qquad}}              % type "\n" to indent at the begin of the line
\newcommand{\red}[1]{{\color{red}#1}}

%% editing comment
\newcommand{\ignore}[1]{}   % ignore this
\newcommand{\cmt}[1]{\begin{sloppypar}\large\textcolor{red}{#1}\end{sloppypar}}

\newcommand{\TODO}[1]{\textcolor{red}{[TODO]\{#1\}}}
\newcommand{\todo}[1]{\textcolor{red}{#1}}
\newcommand{\torevise}[1]{\textcolor{blue}{#1}}
\newcommand{\revise}[1]{\textcolor{blue}{#1}}
\newcommand{\copied}[1]{ \textcolor{red}{[COPIED: #1]}}
\newcommand{\cyy}[1]{\textcolor{green}{#1}}

\newcommand{\comment}[1]{\textcolor{green}{COMMENT: #1}}

\newcommand{\best}[1]{{\underline{\bf{#1}}}}
\newcommand{\second}[1]{\bf{{#1}}}

% self-define variable
\newcommand{\stcganExt}{ST{\text -}CGAN{\text -}BG}
\newcommand{\fullMdl}{BEDSR{\text -}Net}
\newcommand{\bgMdl}{GBCE{\text -}Net}
\newcommand{\srMdl}{SR{\text -}Net}

\def \CVPR{Proceedings of the IEEE Conference on Computer Vision and Pattern Recognition}
\def \ICCP{Proceedings of the IEEE International Conference on Computational Photography}
\def \ICCV{Proceedings of the IEEE International Conference on Computer Vision}
\def \TOG{ACM Transactions on Graphics}
\def \PAMI{IEEE Transactions on Pattern Analysis and Machine Intelligence}
\def \NIPS{Proceedings of the Neural Information Processing Systems Conference}

\begin{abstract}
Modern cameras have limited dynamic ranges and often produce images with saturated or dark regions using a single exposure. Although the problem could be addressed by taking multiple images with different exposures, exposure fusion methods need to deal with ghosting artifacts and detail loss caused by camera motion or moving objects. This paper proposes a deep network for exposure fusion. For reducing the potential ghosting problem, our network only takes two images, an underexposed image and an overexposed one. Our network integrates together homography estimation for compensating camera motion, attention mechanism for correcting remaining misalignment and moving pixels, and adversarial learning for alleviating other remaining artifacts. Experiments on real-world photos taken using handheld mobile phones show that the proposed method can generate high-quality images with faithful detail and vivid color rendition in both dark and bright areas.
\end{abstract}

%Camera motion and moving objects are key challenges in multi-exposure image fusion (MEF) and high dynamic range (HDR) imaging. Previous MEF methods generate unnatural results for difficult scenarios or the lack of input images. Methods recovering HDR images usually register exposure stacks using global alignment methods or optical flows before merging them. Nevertheless, they fail to deal with artifacts caused by misalignment or moving objects. To address these issues, we propose a CNN-based MEF method that produces images preserving color and structural information of exposure stacks even in difficult conditions. Our method only requires 2 low dynamic range (LDR) images to reduce the computational effort. Additionally, it prevents ghosting artifacts by homography estimation and an attention mechanism. Experiments on existing datasets and photos taken on mobile phones show that the proposed method can achieve results of higher human preference.
\section{INTRODUCTION}
\label{sec:introduction}

A camera is an imperfect device for measuring the radiance distribution of a scene because it cannot capture the full spectral content and dynamic range. High dynamic range (HDR) imaging and multi-exposure image fusion (MEF) are techniques to expand the low dynamic range (LDR) due to limitations of the camera sensor. Thanks to the development of imaging devices, we are able to capture a sequence of multi-exposure images in a short time to fulfil the dynamic range of a scene. With the sequence, HDR imaging methods try to recover the response curve and construct an HDR radiance map. MEF methods are applied to blend well-exposed regions from multiple images with different exposures to produce a single visually pleasing LDR image that appears to possess a higher dynamic range. Most MEF or HDR methods perform well with perfectly static sequences~\cite{mertens2007exposure, ma2017robust, prabhakar2017deepfuse}. However, in practice, a certain amount of camera and object motions are inevitable, resulting in ghosting and blurry artifacts in the fused images.

\begin{figure}[htb]
    \centering
    \includegraphics[width=0.48\textwidth]{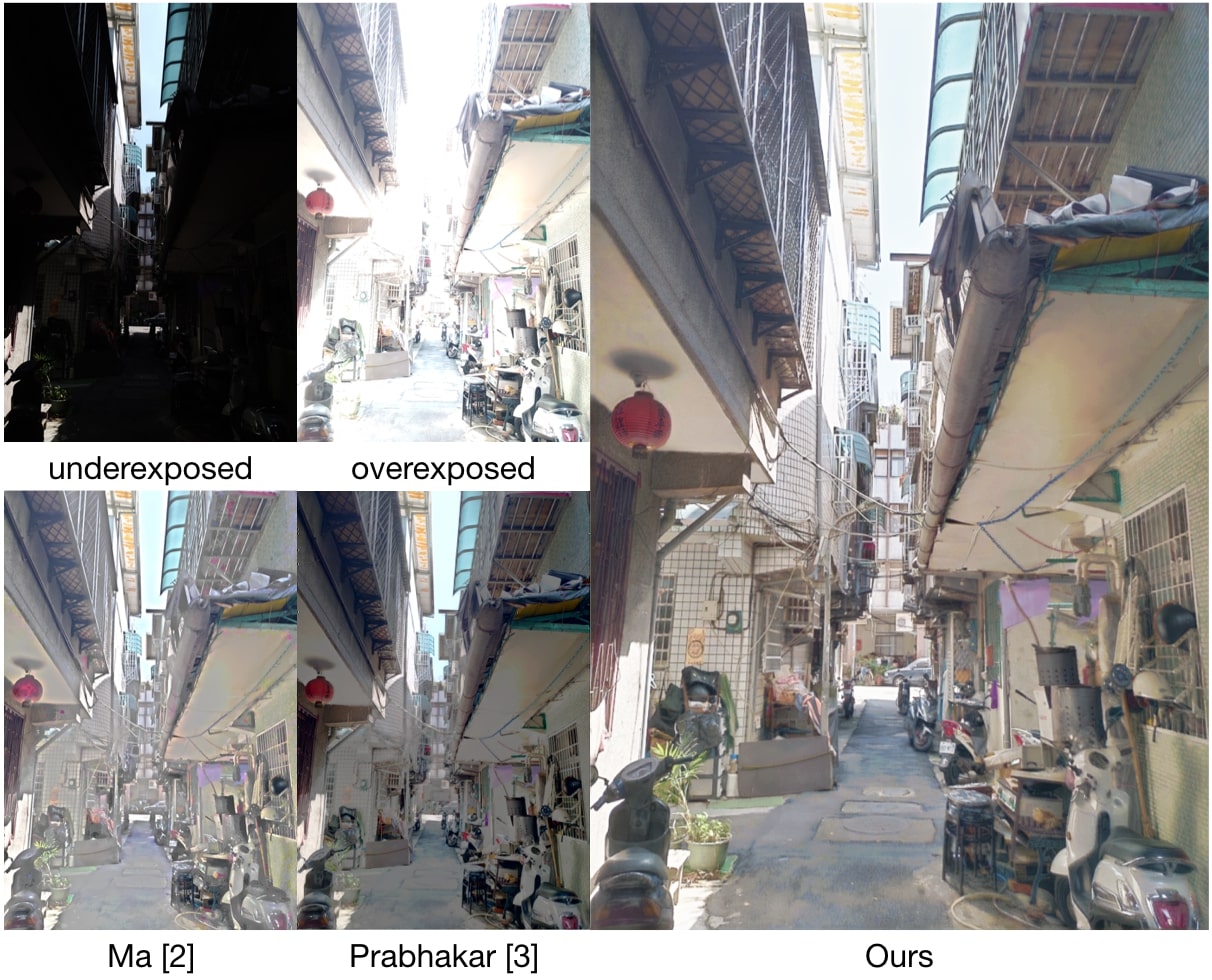}
    \caption{An example result of the proposed method. We propose a learning-based approach to produce a well exposed LDR image given two differently exposed LDR images of a dynamic scene taken via a handheld mobile phone. We also compare our results with two previous methods by Ma~\etal~\cite{ma2017robust} and Prabhakar~\etal~\cite{prabhakar2017deepfuse}.}
    \label{fig:introduction}
\end{figure}

Some techniques have been proposed for addressing the ghosting problem. While camera motion can be compensated using global alignment methods such as median threshold bitmap (MTB)~\cite{ward2003fast}, deghosting of moving objects is more difficult. Global image misalignment can also be compensated by homography~\cite{szpak2015robust}. Wu~\etal~\cite{wu2018deep} applied feature-based registration for estimating homography of multi-exposure images. However, due to parallax in saturated regions, the method cannot produce perfect alignment, and the final output may be blurry. Kalantari~\etal~\cite{kalantari2017deep} aligned exposure stacks using a traditional optical flow technique. Their method corrects the distortion owing to optical flow warping with a CNN, but still have artifacts for extreme dynamic scenes. Yan~\etal~\cite{yan2019attention} proposed an attention-guided network for ghost-free HDR imaging. However, their method requires a large amount of data with moving objects, and needs at least three images in the exposure stack for generating an HDR image.

This paper proposes a deep network for exposure fusion. For reducing potential ghosting, our network is retrained to take only two images, an underexposed image and an overexposed image. The network consists of three main components: a homography estimation network for compensating camera motion, a fusion network with attention learning for reducing misalignment and moving pixels, and a discriminator network for adversarial learning, which alleviates remaining artifacts. 
Our contribution are as follows.
\begin{itemize}[leftmargin=10pt]
    \setlength{\itemsep}{0pt}
    \setlength{\parsep}{0pt}
    \setlength{\parskip}{0pt}
    \item We propose the first deep network for estimating a homography between two differently exposed images and integrate it into an exposure fusion network. Previous methods are either designed for images with similar exposures or based on conventional methods.
    \item We present a model that produces a ghost-free image from two images with large exposure difference via attention learning.
    \item To the best of our knowledge, we are the first to apply adversarial learning for exposure fusion.
\end{itemize}

%our network consists of multiple components. Taking an underexposed image and an overexposed image as input, we first registered the image pair using a effective homography network. Then, the registered images are passed to an attention network to extract features from the image pair. Finally, a merging network merges the features to generate ghost-free fused images. Besides, we utilize a conditional GAN to force the fused images toward the domain of human-preferred reference images. To sum up, our contribution are as follows.
%\begin{itemize}
%    \setlength{\itemsep}{0pt}
%    \setlength{\parsep}{0pt}
%    \setlength{\parskip}{0pt}
%    \item \revise{Our work is the first method applying deep learning homography estimation for different exposure images.}
%    \item \revise{We present a model that produce ghost-free images even the exposure differences of images are large via attention learning.
%    \item To our knowledge, we propose the first application of adversarial learning for exposure fusion.}
%\end{itemize}

\section{METHOD}
\label{sec:method}

Given an underexposed image $I_1$ and an overexposed image $I_2$ of a scene, our goal is to generate an image with better color and structure rendition in both dark and bright regions. Since nowadays photos are often taken with handheld cameras, it is necessary to handle both camera and scene motion. Our method has two stages: alignment and fusion. During alignment, the overexposed image $I_2$ is registered to the underexposed image $I_1$. The aligned images are then blended into a visually pleasing image at the fusion stage.

\begin{figure}[t]
    \centering
    \includegraphics[width=0.48\textwidth]{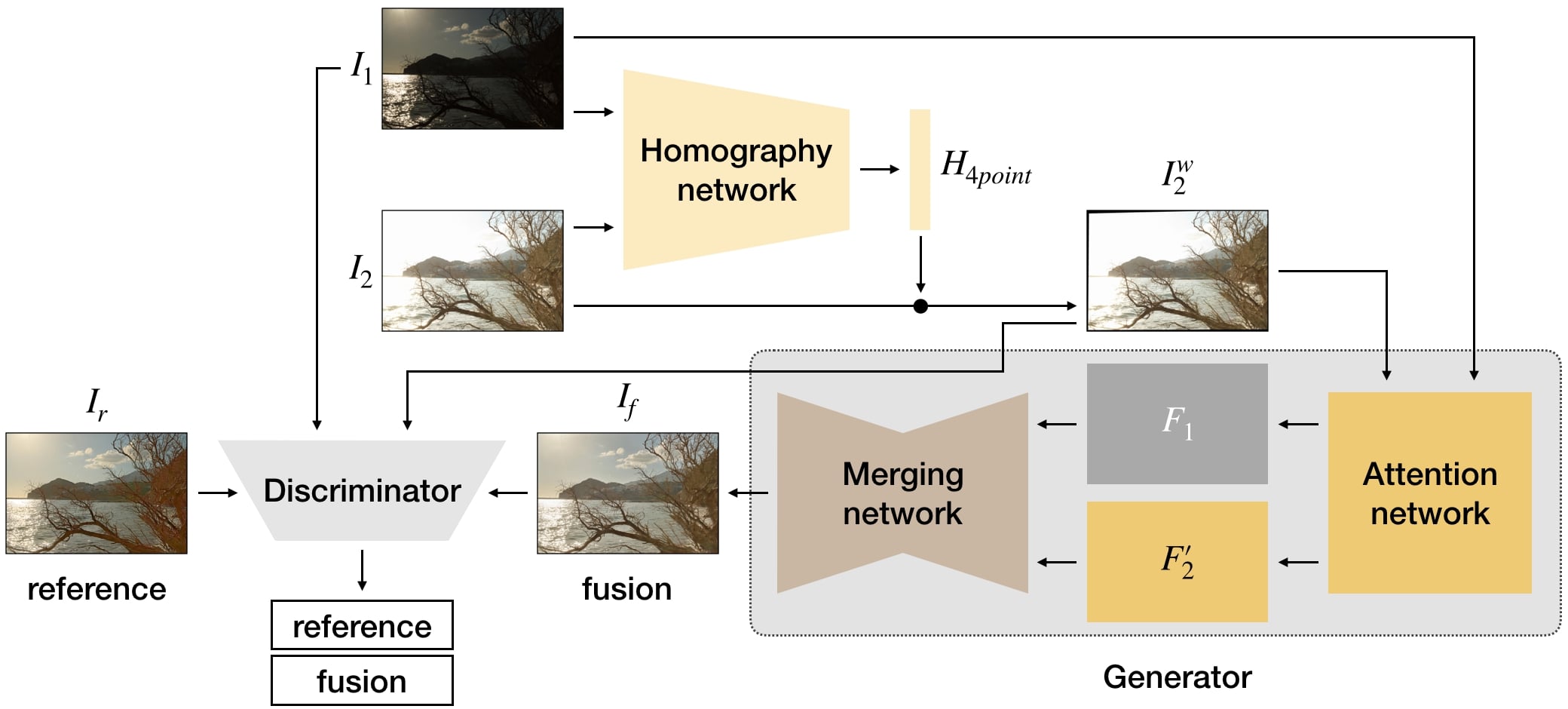}
    \caption{The overview of the proposed method.}
    \label{fig:overview}
\end{figure}

Our method addresses both alignment and fusion problems using a convolutional neural network (CNN). \figref{overview} gives an overview of our method. Given the underexposed image $I_1$ and the overexposed image $I_2$ as the input, first, the homography network estimates the homography that warps $I_2$ to align with $I_1$. Next, the warped overexposed image $I_2^w$ and the underexposed image $I_1$ are fed to the generator to produce an image with better visual quality. Finally, the discriminator takes the underexposed image $I_1$, the warped overexposed image $I_2^w$, and the reference image $I_r$ or the generated result $I_f$ as input and predicts if the third image is the reference image or the generated image. This way, the generator is trained to produce images indistinguishable with the reference images, further improving the visual quality. 
%We will describe the sub-networks below. 

For training the network, we use the SICE dataset collected by Cai~\etal~\cite{cai2018learning}. The SICE dataset contains over 500 exposure stacks of static scenes with high-quality reference images. The reference images were selected after pairwise comparisons of results generated by 13 state-of-the-art multi-exposure fusion or stack-based HDR methods at that time by 13 amateur photographers and 5 volunteers.

%Because there is no ground truth for exposure fusion, and it is necessary for a supervised learning algorithm to have reference images as a learning target. Cai et al. \cite{cai2018learning} proposed a dataset contains over 500 exposure stacks and high quality reference images of static scenes. The reference images were selected after pairwise comparisons of results generated by 13 state-of-the-art multi-exposure fusion or stack-based HDR methods at that time by 13 amateur photographers and 5 volunteers.

\subsection{Homography network}
{\noindent\bf 4-point homography parameterization.} A homography is often represented as a $3 \times 3$ matrix $H_{matrix}$ which maps a pixel $[u, v]$ of the source image to the pixel $[u', v']$ of the destination image by matrix multiplication so that they are aligned. 
%(see Eq. \ref{eq:hmatrix}).
%\begin{equation}
%    \label{eq:hmatrix}
%    \begin{pmatrix}
%        u' \\
%        v' \\
%        1
%    \end{pmatrix}
%    \sim
%    \begin{pmatrix}
%        H_{11} & H_{12} & H_{13} \\
%        H_{21} & H_{22} & H_{23} \\
%        H_{31} & H_{32} & H_{33}
%    \end{pmatrix}
%    \begin{pmatrix}
%        u \\
%        v \\
%        1
%    \end{pmatrix}
%\end{equation}
However, as pointed out by DeTone~\etal~\cite{detone2016deep}, it is difficult to balance the rotational and translation terms by using such a representation in an optimization problem. They suggest to use an alternative 4-point parameterization $H_{4point}$  by using the offsets of the four corner locations, $\Delta u_i = u_i' - u_i, \Delta v_i = v_i' - v_i$ for $i = 1, ..., 4$. We adopt this representation because it leads to more stable optimization.

%They found that an alternative parameterization, one based on a single kind of location variable, namely the corner location, is more suitable for the deep homography estimation task. Letting $\Delta u_1 = u_1' - u_1$ be the u-offset for the first corner, the 4-point parameterization represents a homography as follows:
%\begin{equation}
%    H_{4point} =
%    \begin{pmatrix}
%        \Delta u_1 & \Delta v_1 \\
%        \Delta u_2 & \Delta v_2 \\
%        \Delta u_3 & \Delta v_3 \\
%        \Delta u_4 & \Delta v_4
%    \end{pmatrix}
%\end{equation}
%There is a 1-to-1 mapping between the $3 \times 3$ matrix representation and the 4-point parameterization since they both have 8 DoF (degree of freedom). Once the displacement of the four corners is known, one can easily convert $H_{4point}$ to $H_{matrix}$. This can be performed in many ways. One can use the normalized Direct Linear Transform (DLT) algorithm \cite{hartley2003multiple}, or the function \texttt{getPerspectiveTransform()} in OpenCV. In our research, we implemented the function using TensorFlow, which obtains the matrix by solving a linear least-squares problem.

{\noindent\bf Training example generation.} Since images in the SICE dataset~\cite{cai2018learning} were captured for static scenes with tripods, the images of a scene are well aligned although with different exposures. For preparing training data for the homography network that predicts $H_{4point}$, we induce a random projective transformation to a pair of aligned images at a time. This way, we have an unlimited number of training examples.

\begin{figure*}[t]
\centering
\begin{tabular}{ccc}
    \includegraphics[width=0.32\textwidth]{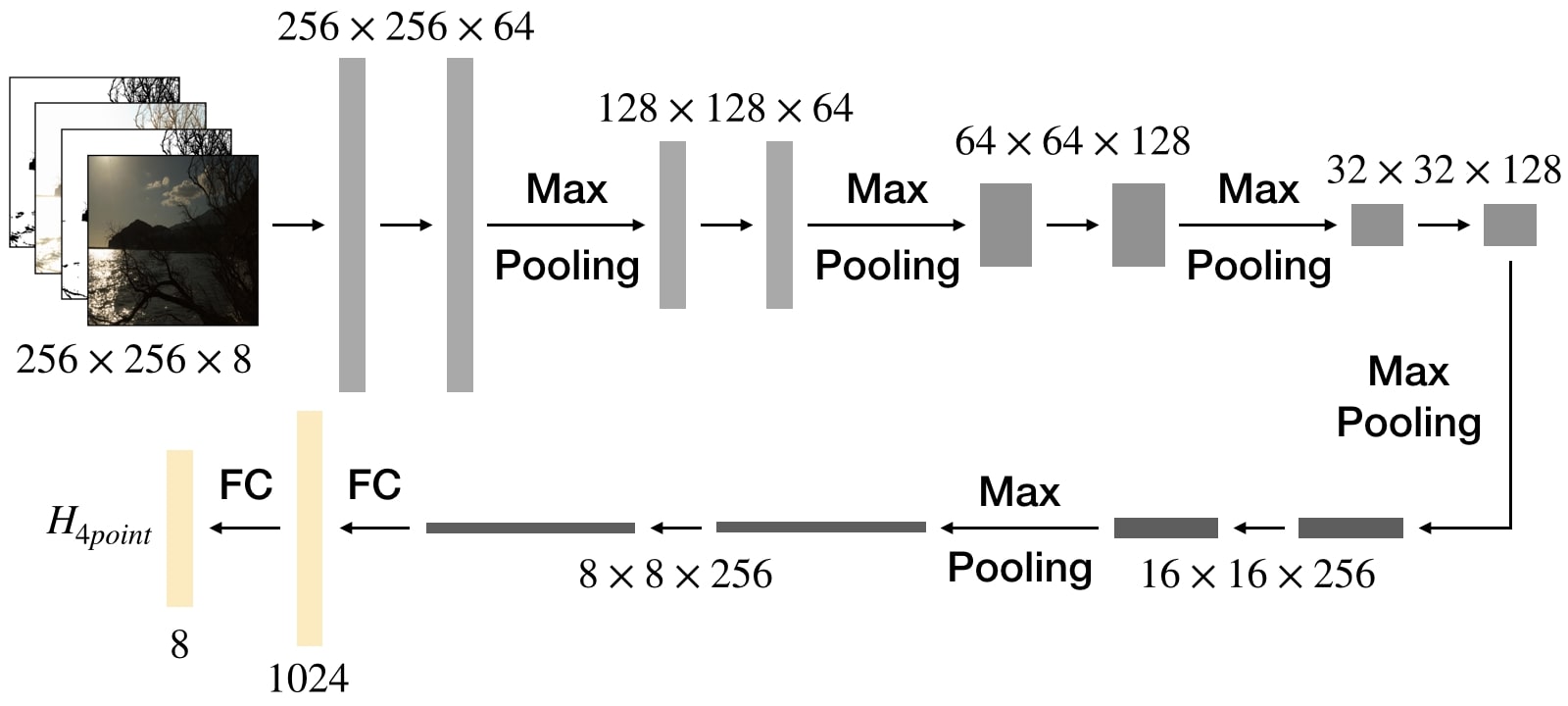} &
    \includegraphics[width=0.35\textwidth]{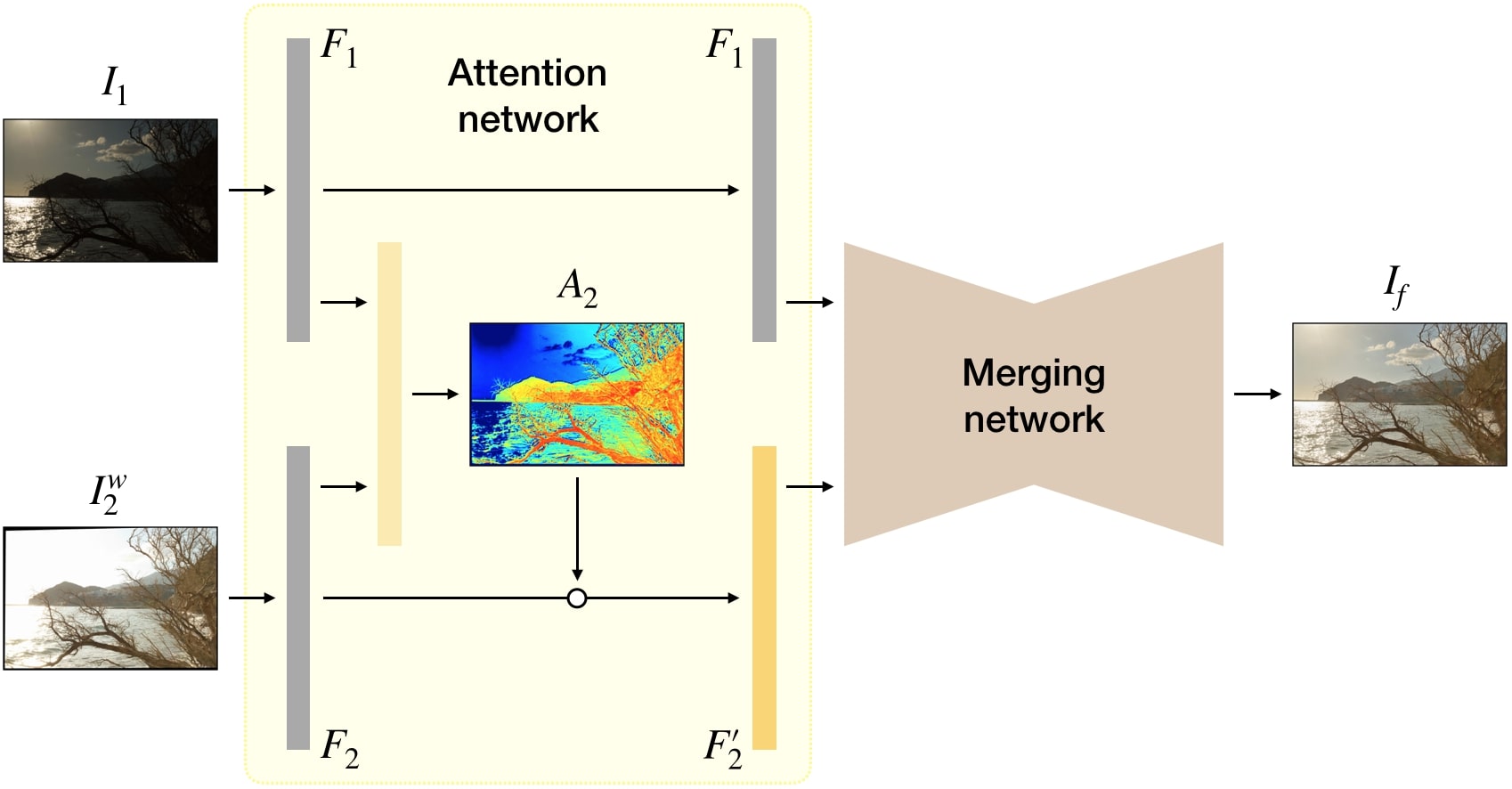} &
    \includegraphics[width=0.30\textwidth]{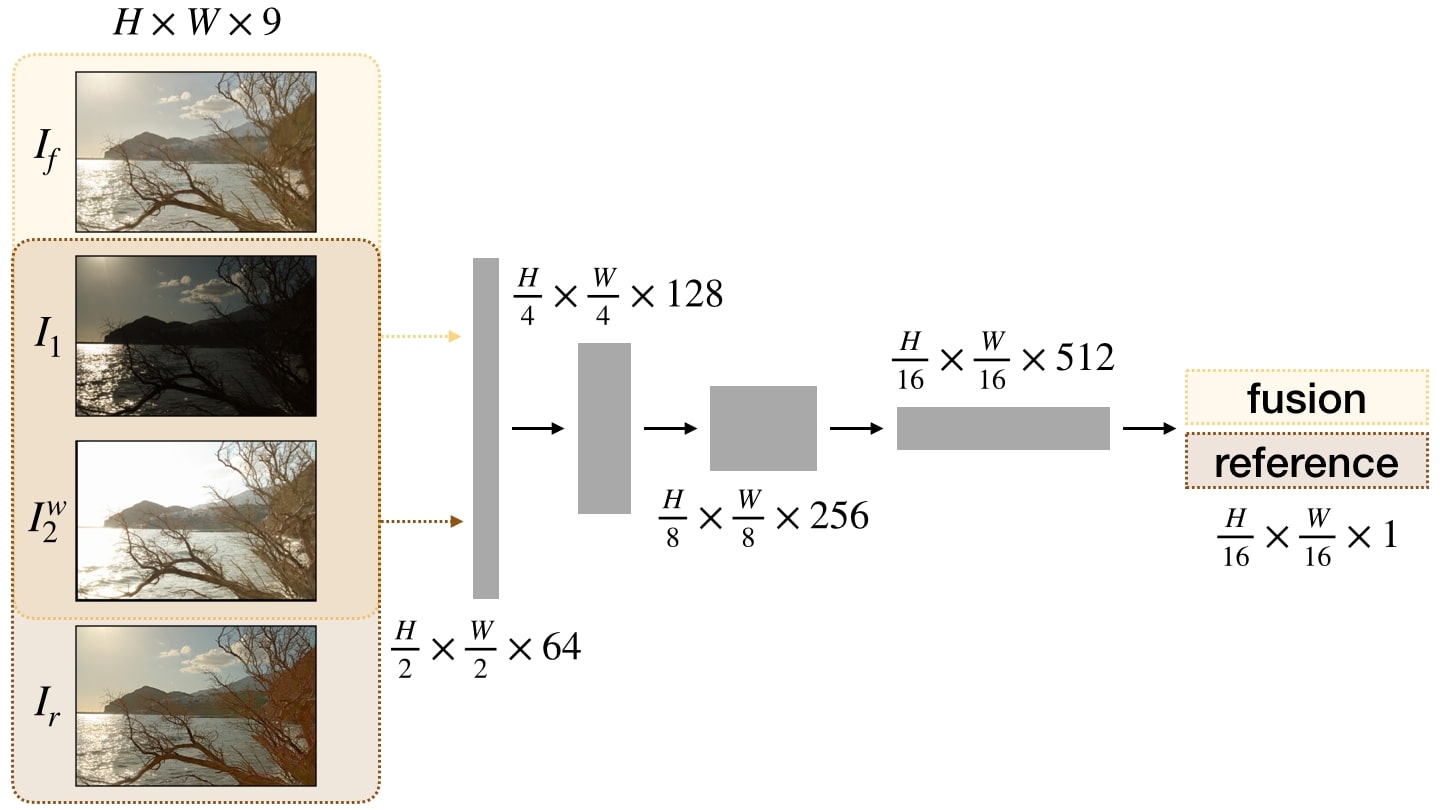} \\
    (a) the homography network & (b) the generator & (c) the discriminator
\end{tabular}
%\vspace{-10pt}
    \caption{The architecture of the sub-networks, the homography network, the generator and the discriminator.}
    \label{fig:modules}
\end{figure*}

{\noindent\bf Homography network $N_h$.} The objective of the alignment stage is to warp the overexposed image $I_2$ to the underexposed image $I_1$. For this purpose, we construct a homography network which is composed of $3 \times 3$ convolutional blocks with instance normalization~\cite{ulyanov2016instance} and leaky ReLUs. The images are resize to $256 \times 256$ as the predicted range of the offsets must be fixed, and the median threshold bitmaps (MTB) of the images are taken as augmented input. The network has 12 convolutional layers with a max pooling layer ($2 \times 2$, stride 2) after every two convolutional layers. The numbers of filters are 64 for the first four layers, 128 for the next four layers, and 256 for the last four layers. At the end, there are two fully connected layers with 1024 and 8 units. %We did not apply Dropout for better performance.
\figref{modules}(a) depicts the architecture of the homography network. The network produces eight real numbers for $H_{4point}$ and is trained using an $L_2$ loss. We then applied the homography for warping $I_2$ into its aligned version, $I_2^w$. Eventually, we have $[H_{4point}, I_2^w] = N_h(I_1, I_2)$.

%First, we resize both images to $256 \times 256$ and train the homography network. Then, we normalized the images to $[-1, 1]$, concatenated them by channel, and sent the $256 \times 256 \times 6$ tensor to the network. Our homography network is composed of $3 \times 3$ convolutional blocks with instance normalization \cite{ulyanov2016instance} and leaky ReLUs, and is architecturally similar to but deeper than the model proposed by DeTone et al. \cite{detone2016deep}. We used 12 convolutional layers with a max pooling layer ($2 \times 2$, stride 2) after every 2 convolutions. The numbers of filters of the 12 layers are 64 for the first 4 layers, 128 for the next 4 layers, and 256 for the last 4 layers. The convolution layers are followed by 2 fully connected layers with 1024 and 8 units. We did not apply Dropout because we wanted to feed the result of the homography network to the fusion phase and it may reduce the alignment performance during training. The network directly produces 8 real-valued $H_{4point}$ and is trained using an $l_2$ loss. As we have the parameters of homography (scaled by the original size of the input image), we can transform the overexposed image $I_2$ to $I_2^w$ to align with the underexposed image $I_1$ (see Eq. \ref{eq:N_h}).
%\begin{equation}
%    \label{eq:N_h}
%    [H_{4point}, I_2^w] = N_h(I_1, I_2).
%\end{equation}

\subsection{Generator}

Given the registered images $I_1$ and $I_2^w$, the generator attempts to fuse them into a visually pleasing image. As shown in \figref{modules}(b), our generator consists of two modules: the attention network and the merging network. 

{\noindent\bf Attention network $N_a$.} Inspired by Yan'~\etal~\cite{yan2019attention}, we use an attention mechanism on the warped overexposed image $I_2^w$ to highlight the well-aligned area for generating ghost-free fusion results. Given $I_1$ and $I_2^w$, the attention network first uses a shared encoding layer to extract feature maps $F_1$ and $F_2$ with 64 channels. $F_1$ and $F_2$ are then sent to the convolutional attention module $a_2(\cdot)$ for obtaining the attention map $A_2$ for $I_2^w$, \ie, $A_2 = a_2(F_1, F_2)$.
%To obtain the attention map for the warped overexposed image, we feed the features $F_1$ and $F_2$ to the convolutional attention module $a_2(\cdot)$, and then acquire the attention map $A_2$ for $I_2^w$ (see Eq. \ref{eq:a2}), which is the same size of the feature map $F_2$ and values of $A_2$ are in the range $[0, 1]$.
%\begin{equation}
%    \label{eq:a2}
%    A_2 = a_2(F_1, F_2).
%\end{equation}
The attention module has two convolutional layers with 64 $3 \times 3$ filters, respectively followed by a leaky ReLU activation and a sigmoid function. 
%As a result, the attention module can use the stacked input feature maps $F_1$ and $F_2$ to obtain the 64-channel attention map $A_2$. 
The predicted attention map is used to attend the features of the warped overexposed image by $F_2' = A_2 \circ F_2$,
%\begin{equation}
%    F_2' = A_2 \circ F_2,
%\end{equation}
where $\circ$ denotes the element-wise multiplication and $F_2'$ is the feature map after attention. As a result, the attention network $N_a$ takes $I_1$ and $I_2^w$ from the homography network and generates $F_1$ and $F_2'$: $[F_1, F_2'] = N_a(I_1, I_2^w)$.
The attention maps can suppress the misaligned and saturated regions in the overexposed image, avoiding the unfavorable features getting into the merging process and alleviating ghosting accordingly.

%\begin{equation}
%    \label{eq:N_a}
%    [F_1, F_2'] = N_a(I_1, I_2^w).
%\end{equation}

%\begin{figure}[htb]
%    \centering
%    \includegraphics[width=0.48\textwidth]{figures/method/generator.jpg}
%    \caption{The architecture of the generator.}
%    \label{fig:generator}
%\end{figure}

{\noindent\bf Merging network $N_m$.} We model the merging network after U-Net~\cite{ronneberger2015u} with seven downsampling convolutional layers followed by seven upsampling deconvolutional layers. Each layer is composed of 64 $4 \times 4$ filters.
%Except the first convolutional layer and the last deconvolutional layer, we apply instance normalization which performs better then batch normalization in our experiments.
By taking inputs from the attention network, $F_1$ and $F_2'$, the merging network produces an image $I_f$, exhibiting abundant color and structure information in both dark and bright regions by $[I_f] = N_m(F_1, F_2')$.
%\begin{equation}
%    \label{eq:N_m}
%    [I_f] = N_m(F_1, F_2').
%\end{equation}

The optimization target of the merging network is to minimize the difference between the fused image $I_f$ and the reference image $I_r$. 
%However, the tradition $L_2$ loss function often leads to information loss of high frequency. Thus, w
For measuring differences, we use the perceptual loss that has been successfully applied to tasks such as image synthesis~\cite{chen2017photographic}, super-resolution~\cite{ledig2017photo}, and reflection removal~\cite{zhang2018single}. We obtain the perceptual loss by feeding $I_f$ and $I_r$ through a pretrained VGG-19 network $\Phi$. We compute the $L_1$ loss between $\Phi(I_f)$ and $\Phi(I_r)$ for the layers `conv1\_2', `conv2\_2', `conv3\_2', `conv4\_2', `conv5\_2', and the `input' layer in VGG-19:
\begin{equation}
    L_{feat} = \sum_{I_1, I_2} \sum_l \lambda_l \parallel \Phi_l(I_r) - \Phi_l(I_f) \parallel_1,
\end{equation}
where $\Phi_l$ represents the $l$-th layer in the VGG-19 network. We assign the weighting terms $\lambda_l$ as $\frac{1}{2.6}$, $\frac{1}{4.8}$, $\frac{1}{3.7}$, $\frac{1}{5.6}$, $\frac{1}{0.15}$ for the convolutional layers and $1$ for the input layer~\cite{zhang2018single}.

\subsection{Discriminator}

The fused image produced by the generator could still suffer from undesirable color degradation and halo effects. For further improving the visual quality, we adopt the idea of adversarial learning by modeling after the conditional PatchGAN~\cite{isola2017image}. We construct our discriminator $N_D$ with five convolutional layers with $4 \times 4$ convolutional blocks. The numbers of filters are 64, 128, 256, 512, and 1. We use instance normalization for the middle three layers. Except the final convolutional layer, the stride number is 2 and the activation function is leaky ReLU. \figref{modules}(c) shows the architecture. The discriminator attempts to discriminate between patches $I_r$ from the reference images and patches $I_f$ generated by $N_m$ conditioned on $I_1$ and $I_2$. With the discriminator, the generator is trained to produce results more similar to the reference.
%The goal is to push the fused image $I_f$ toward the domain of the human-preferred images generated by state-of-the-art algorithms.

%\begin{figure}[htb]
%    \centering
%    \includegraphics[width=0.48\textwidth]{figures/method/discriminator.jpg}
%    \caption{The architecture of the discriminator.}
%    \label{fig:discriminator}
%\end{figure}

The adversarial loss is defined as LSGAN~\cite{mao2017least}. The optimization goal of the discriminator $N_D$ is:
\begin{equation}
    \mathop{\arg\min}_{N_D} \sum_{I_1, I_2} \frac{1}{2} [N_D(I_1, I_2, I_f)^2 + (N_D(I_1, I_2, I_r) - 1)^2],
\end{equation}
where $N_D(I_1, I_2, x)$ outputs the probability that $x$ is a high quality fused image given the input images $I_1$, $I_2$. In addition, our adversarial loss $L_{adv}$ is:
\begin{equation}
    L_{adv} = \sum_{I_1, I_2} (N_D(I_1, I_2, I_f) - 1)^2.
\end{equation}
The goal of the generator is to minimize $L_{feat}$ and $L_{adv}$:
\begin{equation}
    \label{eq:L_G}
    \mathop{\arg\min}_{N_a, N_m} \sum_{I_1, I_2} w_1 L_{feat} + w_2 L_{adv},
\end{equation}
where $w_1\!\!=\!\!1$ and $w_2\!\!=\!\!0.01$ are parameters balancing losses. 

\begin{figure*}[t]
    \centering
    \includegraphics[width=0.84\textwidth]{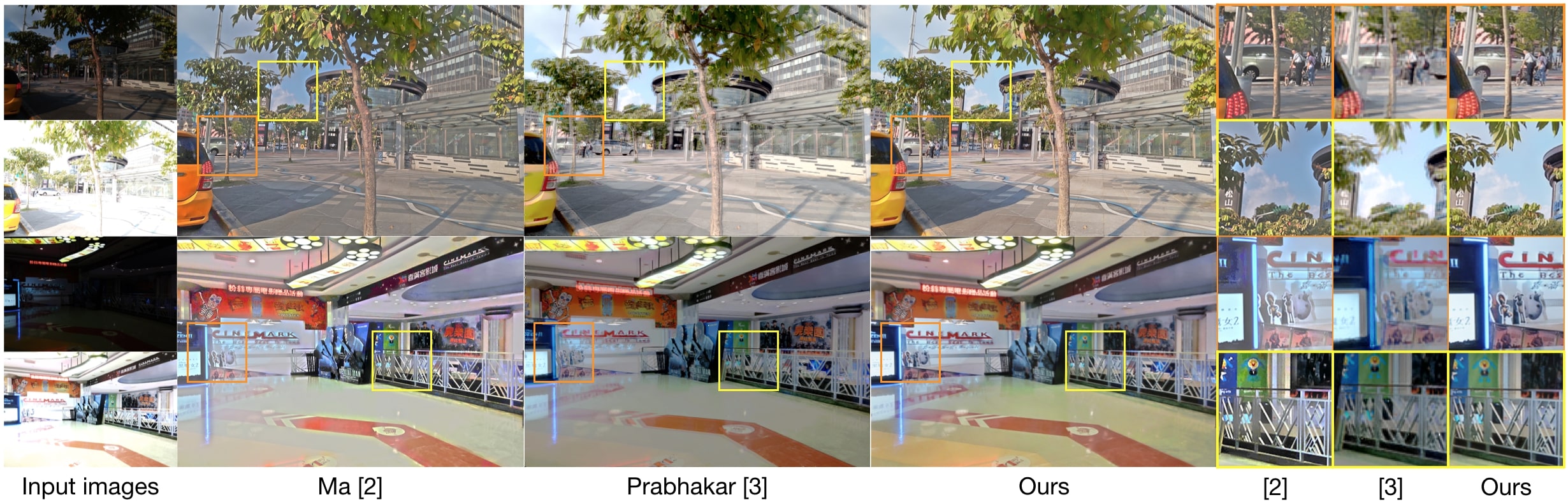}
    \vspace{-5pt}
    \caption{Comparison with Ma~\etal~\cite{ma2017robust} and DeepFuse~\cite{prabhakar2017deepfuse}.}
    \label{fig:exposure_fusion}
\end{figure*}

\section{EXPERIMENTS}
\label{sec:experiments}

%As we need a stable homography estimation before merging the exposure pairs, 
We first train our homography network $N_h$ for 500 epochs, and then train the remaining networks ($N_D$, $N_a$, $N_m$) for 200 epochs. The whole training process takes about 27 hours on a single P100 GPU. We use a batch size of 4 patches and each epoch has 471 iterations. We optimize our networks using Adam optimizer~\cite{kingma2014adam} with a learning rate of 0.0001.

\begin{figure}[t]
    \centering
    \includegraphics[width=0.48\textwidth]{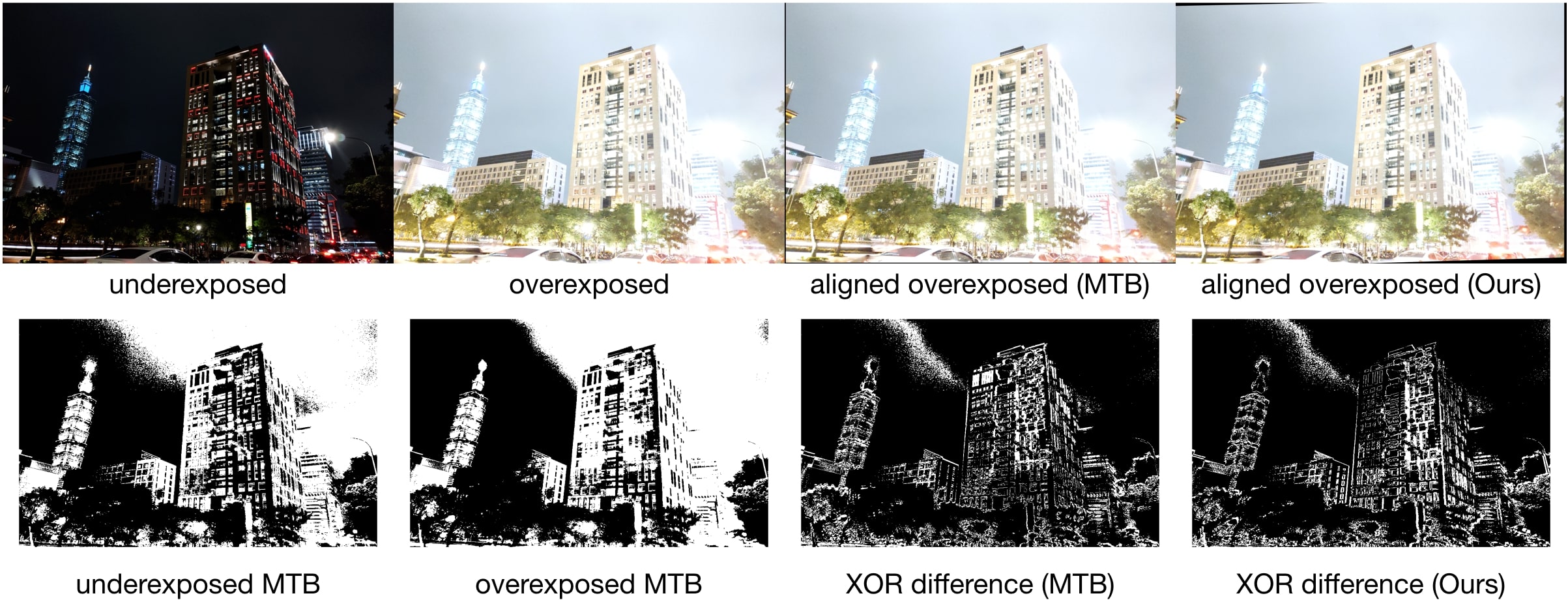}
    \caption{Comparison of homography estimation with MTB.}
    \label{fig:homography_estimation}
\end{figure}

For training, we use the training set of SICE dataset by Cai~\etal~\cite{cai2018learning}. All images are resized to $1200 \times 800$.
%with the function \texttt{resize()} using \texttt{INTER\_AREA} interpolation method in OpenCV. 
Since our approach requires two differently exposed images as input, we sort all exposure stacks in the dataset by luminance, and take the $i$-th and the $N-i+1$-th image in an exposure stack which contains $N$ images as a training example, where $i = 1, ..., \frac{N}{2}$. 
We generate image patches of the size $256 \times 256$ in run time related by a 4-point homography parameterization randomly sampled by \texttt{truncated\_normal()} in TensorFlow. 
We perturbed each corner of the $256 \times 256$ patch by a maximum of one eighth of the image width. We avoid larger random disturbance for avoiding extreme transformation which might affect the quality of fusion.
We use the testing set of SICE dataset as the validation set to monitor overfitting. As for the testing set, we test our method on photos taken using a handheld mobile phone with the exposure bracketing mode. 
%We use Exposure Bracketing mode of an Android app \textit{Open Camera}~\cite{opencamera} to capture three images with different exposures sequentially, and the darkest image and the brightest image are used as the input for testing. 
To test the robustness of our approach, the exposure values are selected automatically by the app.

{\noindent \bf Homography estimation.} \figref{homography_estimation} compares our homography network with MTB~\cite{ward2003fast}. 
%Removing a border of 30 to avoid outliers in the edges of the warped image, we compute the XOR difference of the MTB of the underexposed and the aligned overexposed image. The results are shown in Figure \ref{fig:homography_estimation}. 
On the bottom right, we show the XOR difference, indicating differences after alignment. Our result has less misaligned pixels than MTB as shown in the XOR difference. 
Our success rate is about 86.63\% for 60 pairs of photos taken by handheld phone cameras. 

{\noindent \bf Exposure fusion.} \figref{exposure_fusion} compares our method with the patch-based approach by Ma~\etal~\cite{ma2017robust} and the DeepFuse approach by Prabhakar~\etal~\cite{prabhakar2017deepfuse}. Note that we do not have ground-truth images here as the images were captured by handheld cameras. The method of Ma~\etal requires more images for generating reasonable results. With two images, their results exhibit color distortion and halo artifacts. DeepFuse requires static scenes and their results are blurry in these examples. 

%\subsection{Ablation studies}

\begin{figure}[t]
    \centering
    \includegraphics[width=0.48\textwidth]{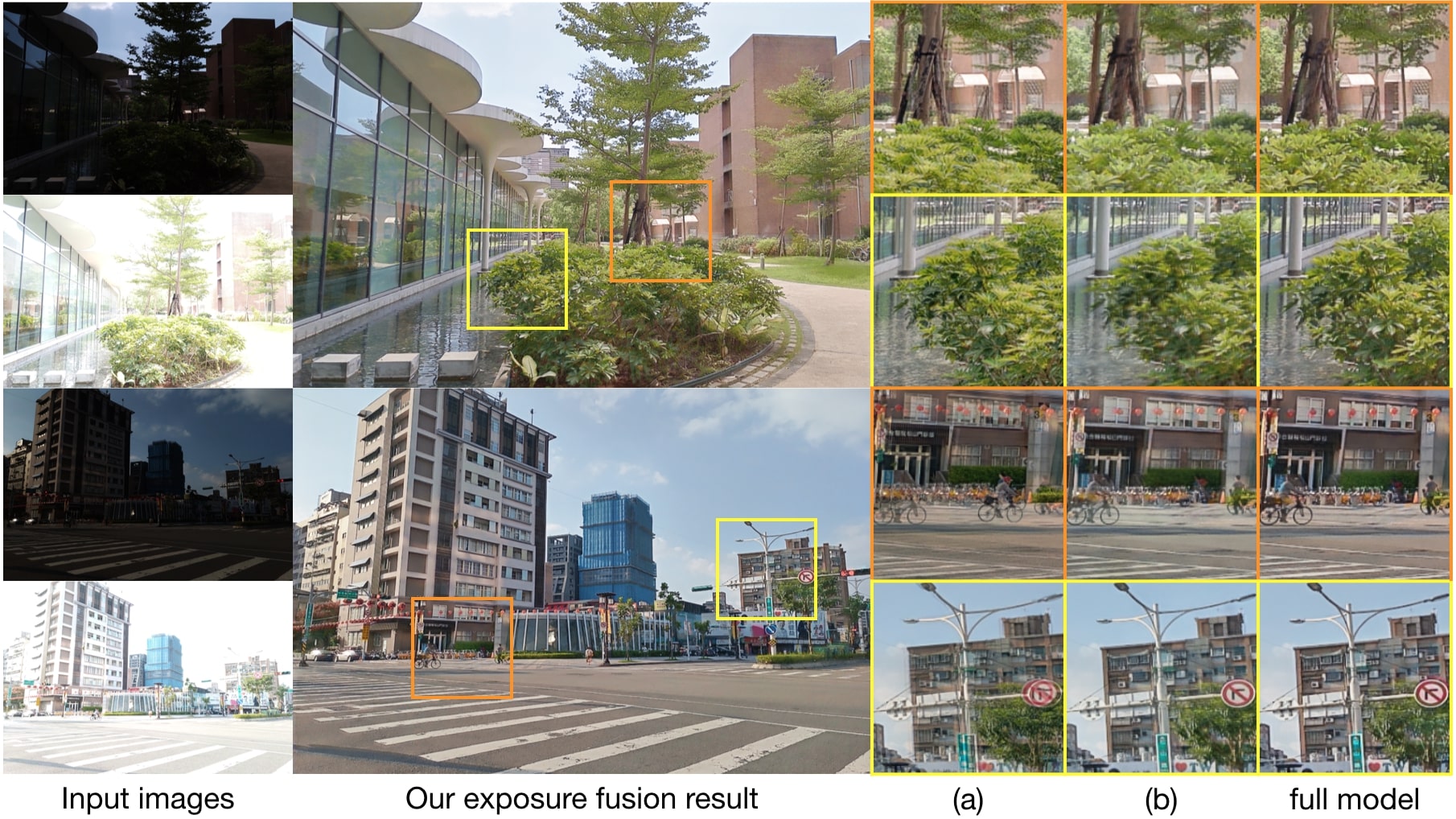}
    \caption{Ablation study. (a) w/o homography network (b) w/o attention mechanism. %(c) l1 loss (d) perceptual loss
    }
    \label{fig:ablation_studies}
\end{figure}

{\noindent\bf Ablation study on the architecture.} For investigating the contribution of individual components, we compare variants of the proposed network, \textit{w/o $N_h$} and \textit{w/o $N_a$}. As shown in \figref{ablation_studies}, the variant \textit{w/o $N_h$} fails to align well when there is camera motion or moving objects, leading to blurry results. Although the \textit{w/o $N_a$} variant can produce sharper results, it still has ghosting artifacts in saturated regions. The full model can alleviate ghosting artifacts and provide sharper details.

%{\noindent\bf Study on the losses.} \revise{In Fig. \ref{fig:ablation_studies}, We also compare the results of our model optimized by different loss functions. We found that the model trained by the $l_1$ loss of the fused results and the reference images produced images with undesirable color degradation and unrealistic halo effects. The model trained by perceptual losses generated more perceptually appealing results but the color is unnatural. The adversarial loss helps the model to generate realistic and attractive fused results.}

\section{CONCLUSIONS}
\label{sec:conclusions}

This paper presents a deep learning method for exposure fusion. For reducing ghosting artifacts and rendering sharp details, our model integrates homography estimation for compensating camera motion, attention mechanism for reducing influence of moving objects and adversarial learning for further improving visual quality. With these modules together, our method can generate images with vivid color and sharp details in both dark and bright areas from a pair of underexposed and overexposed images.

%The application of multi-exposure image fusion on mobile phones has been limited due to the requirement of several different exposed images and ghosting artifacts caused by camera motion and moving objects. We proposed a learning-based approach of combining homography estimation and exposure fusion which requires only a differently exposed image pair to generate a pleasing well exposed image. Our method for homography estimation can be adopted as an alternative of extensively used MTB alignment, and the employment of attention mechanism, perceptual losses, and adversarial learning offers the prospect of more applications in multiple images fusion.

% References should be produced using the bibtex program from suitable
% BiBTeX files (here: strings, refs, manuals). The IEEEbib.bst bibliography
% style file from IEEE produces unsorted bibliography list.
% -------------------------------------------------------------------------
\bibliographystyle{IEEEbib}
\bibliography{chen}

\end{document}